# On-Demand Coherent Nanolaser Metalens and Beam Steering Enabled by Physics-Informed Neural Networks

Omar A. M. Abdelraouf [1,*]

**Affiliations:**
[1]Institute of Materials Research and Engineering, Agency for Science, Technology, and Research (A*STAR), 2 Fusionopolis Way, #08-03, Innovis, Singapore 138634, Singapore

*Corresponding author: Omar A.M. Abdelraouf (email: Omar_Abdelrahman@a-star.edu.sg)

## ABSTRACT

The integration of artificial intelligence with physical modeling offers a transformative route for accelerating the design of active nanophotonic devices. Here, we present NanoPhotoNet-Lase, a physics-informed neural network (PINN) framework that embeds the electromagnetic and rate equations of lasing directly into its learning process to expedite the design of metasurface nanolasers. By coupling Maxwell's vector Helmholtz equation with the four-level population dynamics of dye gain media, the model achieves physics-guided prediction of optical responses, enabling rapid estimation of lasing thresholds across arbitrary nanostructure geometries and material configurations. Using high-index metasurfaces cavity, the NanoPhotoNet-Lase model identifies optimized geometries supporting quasi–bound states in the continuum (BICs) with strong confinement and high-quality factors. The predicted lasing was experimentally realized using Rhodamine B dye as gain medium. The measured lasing threshold ($P_{th}$~565 μJ/cm$^2$) and emission wavelength (λ~620 nm) exhibited <1% deviation from model predictions. Importantly, the framework enables design phase-gradient nanolaser metalens and beam steering that demonstrated coherent, directional, focused or steered emission. This work bridges physics-informed machine learning with experimental nanophotonics, establishing a scalable paradigm for real-time, physically interpretable design of coherent light-emitting metasurfaces.

**KEYWORDs**: NanoPhotoNet-Lase; Physics-informed neural networks (PINNs); Metasurface nanolaser; Bound states in the continuum (BIC); Niobium pentoxide ($Nb_2O_5$); Rhodamine B gain medium; Active metasurfaces; Phase-gradient metalens; Coherent beam steering; AI-assisted photonic design

## INTRODUCTION

The sustained drive toward miniaturization across photonics and optoelectronics has catalyzed an intense focus on nanoscale coherent light sources, or nanolasers, which combine strong spatial confinement with coherent emission at subwavelength dimensions.[1-9] These devices underpin emerging technologies spanning on-chip optical communication, quantum information processing, high-resolution bio-imaging, and coherent light field modulation.[10-15] Among various nanolaser architectures, metasurface nanolaser leveraging bound states in the continuum (BICs) have emerged as especially promising due to their ultra-high quality factors (Q-factor), reduced radiation loss, and robust field localization within compact resonant cavities. These BIC-based platforms provide a path toward ultrathin, integrable coherent light sources with engineered emission properties.[16-21]

Despite this potential, the design of BIC nanolasers remains fundamentally limited by the high-dimensional complexity of their parameter space.[22] The modal characteristics and lasing thresholds depend sensitively on an intertwined set of variables: geometric parameters of nanostructured resonators, refractive index distributions, gain–cavity coupling efficiencies, and the nonlinear carrier dynamics of the active medium. Conventional numerical strategies such as full three-dimensional Finite-Difference Time-Domain (FDTD) or Finite Element Method (FEM) simulations have proven invaluable for exploring these devices, yet the computational demands are severe. Each simulation involving coupled Maxwell–Bloch or laser rate equations can require hours of high-performance computing, rendering large-scale sweeps, stochastic searches, or inverse design workflows practically infeasible.[23] The prohibitive computational latency thus bottlenecks innovation in nanolaser discovery, in particular when active and resonant degrees of freedom must be co-optimized simultaneously.

In recent years, artificial intelligence (AI) and machine learning (ML) have revolutionized optical design, offering data-driven acceleration of metasurface and photonic component optimization.[24-27] Neural networks, Gaussian processes, and variational autoencoders have been employed to perform near-instantaneous predictions of far-field responses or to infer geometry–spectra relationships from precomputed datasets.[28-30] However, such purely data-driven approaches are inherently *black-box* and often lack *physical interpretability and generalization*. Their accuracy deteriorates outside the training space or when extrapolating to new materials, and they typically require tens of thousands of labeled electromagnetic simulations, which is an unviable cost for systems as complex as nanolasers.[31] Furthermore, these networks fail to enforce physical constraints such as energy conservation, population inversion, or radiation boundary conditions, leading to nonphysical predictions when applied to gain media or resonant near-field problems.

Physics-Informed Neural Networks (PINNs) have recently emerged as an elegant solution to this limitation by embedding the governing partial differential equations (PDEs) directly into the learning process.[32-34] Initially developed for fluid mechanics, solid mechanics, and quantum systems, PINNs are now gaining traction in photonics for tasks such as solving the Helmholtz equation, modeling nonlinear propagation, and inverse design of resonant cavities. Unlike conventional neural networks trained on labeled datasets, PINNs minimize the residuals of PDEs, where Maxwell's equations and the rate equations underlying laser kinetics, ensuring that the learned representations are intrinsically compliant with physics.[35-37] This fusion of *deep learning with physical laws* enables data-efficient training and extrapolation to unobserved regimes. However, despite rapid progress in applying PINNs to passive photonic resonance problems, their integration with active gain dynamics and lasing physics remains largely unexplored.

To address this gap, we propose NanoPhotoNet-Lase, a unified physics-informed neural network architecture that directly couples the electromagnetic response of metasurfaces with the carrier population kinetics of an organic gain medium. The framework embeds the vector Helmholtz equation and four-level atomic rate equations into its composite loss function, allowing the network to self-consistently predict lasing threshold and field distributions given arbitrary geometric and material parameters. NanoPhotoNet-Lase ingests unstructured spatial data which represents metasurface meta-atoms of variable dimensions and refractive indices and derives the corresponding local electric field and polarization distributions while maintaining permutation invariance and physical consistency. The model thus bypasses the need for extensive training on labeled datasets while preserving interpretability by enforcing energy and population inversion constraints at every iteration.

We deploy this framework for the accelerated design of Niobium Pentoxide ($Nb_2O_5$) metasurface nanolasers doped with Rhodamine B (Rh B). $Nb_2O_5$ was chosen for its combination of high refractive index ($n$ ~ 2.3 in the visible), low optical loss, and compatibility with atomic layer deposition (ALD),[38] offering strong confinement of quasi-BIC resonant modes. RhB acts as the gain medium, harnessed through its high fluorescence quantum yield, tunability in the visible spectrum, and compatibility with solid-state PVA host matrices. The physics-informed model rapidly identifies configurations of nanopillar radius, height, and pitch that promote high-Q quasi-BIC modes with optimal gain overlap. Compared to conventional FDTD workflows, NanoPhotoNet-Lase achieves over $\times 10^5$ acceleration in predicting optical modes and lasing thresholds while preserving quantitative accuracy in forward design. Fabricated $Nb_2O_5$:RhB metasurface nanolasers experimentally validate the predictions of the NanoPhotoNet-Lase model, showing excellent agreement in both lasing wavelength and threshold fluence. Moreover, we leverage PINN's forward predictive capability to design phase-gradient nanolaser and beam steering arrays that coherently focus and steer the emitted PL according to generalized Snell's law. These demonstrations showcase the versatility of the proposed framework, merging machine learning, laser physics, and nanofabrication into a cohesive "lab-on-algorithm" paradigm. This work thus introduces a physics-informed AI paradigm for active nanophotonics, where the physical laws of electromagnetism and quantum laser dynamics guide a differentiable learning process. Beyond accelerating laser design, this approach offers an interpretable, data-efficient, and experimentally validated pathway toward fully programmable coherent metasurfaces and the next generation of AI-assisted photonic devices.

## RESULTS AND DISCUSSION

### Physics-Informed Design Concept and the NanoPhotoNet-Lase Architecture

The realization of coherent metasurface nanolasers requires the simultaneous, rigorous treatment of two strongly coupled physical domains: the electromagnetic resonant response of subwavelength nanostructures and the nonequilibrium carrier dynamics of the active gain medium. In conventional numerical frameworks, these processes are typically solved independently through iterative coupling between Maxwell solvers and rate-equation formalisms. This traditional approach results in substantial computational complexity, rendering large-scale parameter sweeps and inverse design practically infeasible. To overcome this fundamental bottleneck, we introduce the NanoPhotoNet-Lase framework, a unified PINN designed to self-consistently predict optical fields, polarization dynamics, and lasing thresholds directly from arbitrary metasurface geometries and material distributions.

Our proposed platform utilizes a dielectric metasurface cavity that supports high-quality-factor (*Q*-factor) quasi- BIC operating in the visible regime. The unit cell consists of two niobium pentoxide ($Nb_2O_5$) rectangular pillars resting on a quartz substrate. $Nb_2O_5$ was strategically selected for its high refractive index ($n \sim 2.3$ in the visible spectrum) and low absorption losses, which are essential for supporting tightly confined quasi-BIC modes. To provide the necessary optical gain, the metasurface cavity are embedded within a solid-state matrix of polyvinyl alcohol (PVA) doped with Rhodamine B (Rd B) dye (Figure 1b). The active cavity is externally pumped by a 532 nm nanosecond laser (ns-laser), which excites the Rd B molecules to emit stimulated photons near the 620 nm wavelength (Figure 1a).

The theoretical formulation underlying the NanoPhotoNet-Lase model is deeply rooted in the coupled solution of Maxwell's equations, Lorentz polarization dynamics, and carrier rate equations. These governing physical laws define the constraints that are explicitly incorporated into the network's composite loss function. Rather than relying on traditional spatial discretization schemes, the passive electromagnetic response of the metasurface is governed by the frequency-domain vector Helmholtz equation (Equation 1 in SI). The NanoPhotoNet-Lase model learns these field solutions by minimizing physics residuals over randomly sampled collocation points across the computational domain.

Simultaneously, the model evaluates the active lasing state by incorporating a four-level laser system appropriate for the Rd B organic dye. The temporal evolution of the carrier populations is dictated by coupled rate equations (Equations 4–6 in SI), which establish a self-consistent feedback loop between the optical confinement and the carrier dynamics. Efficient lasing necessitates rapid depopulation of the lower energy level to sustain a continuous population inversion (Equation 7 in SI). The AI architecture consists of 3D spatial point clouds where each point contains coordinates, desired electric field profile, and local refractive index data (Figure 1c). Shared multilayer perceptrons (MLPs) extract local geometric features, while symmetric max-pooling aggregates global structural information, producing a latent descriptor that accurately encodes periodicity and resonant coupling strength. The total loss function driving the backpropagation integrates the electromagnetic residual loss, the carrier-dynamics rate equations, a specific gain-threshold loss, and a sparse supervised field data loss to prevent convergence toward trivial vacuum solutions. By actively enforcing these physics-informed penalties, the AI framework bypasses the computational bottlenecks of traditional solvers and directly converges on fixed-point solutions of the nonlinear coupled Maxwell-Bloch system within milliseconds.

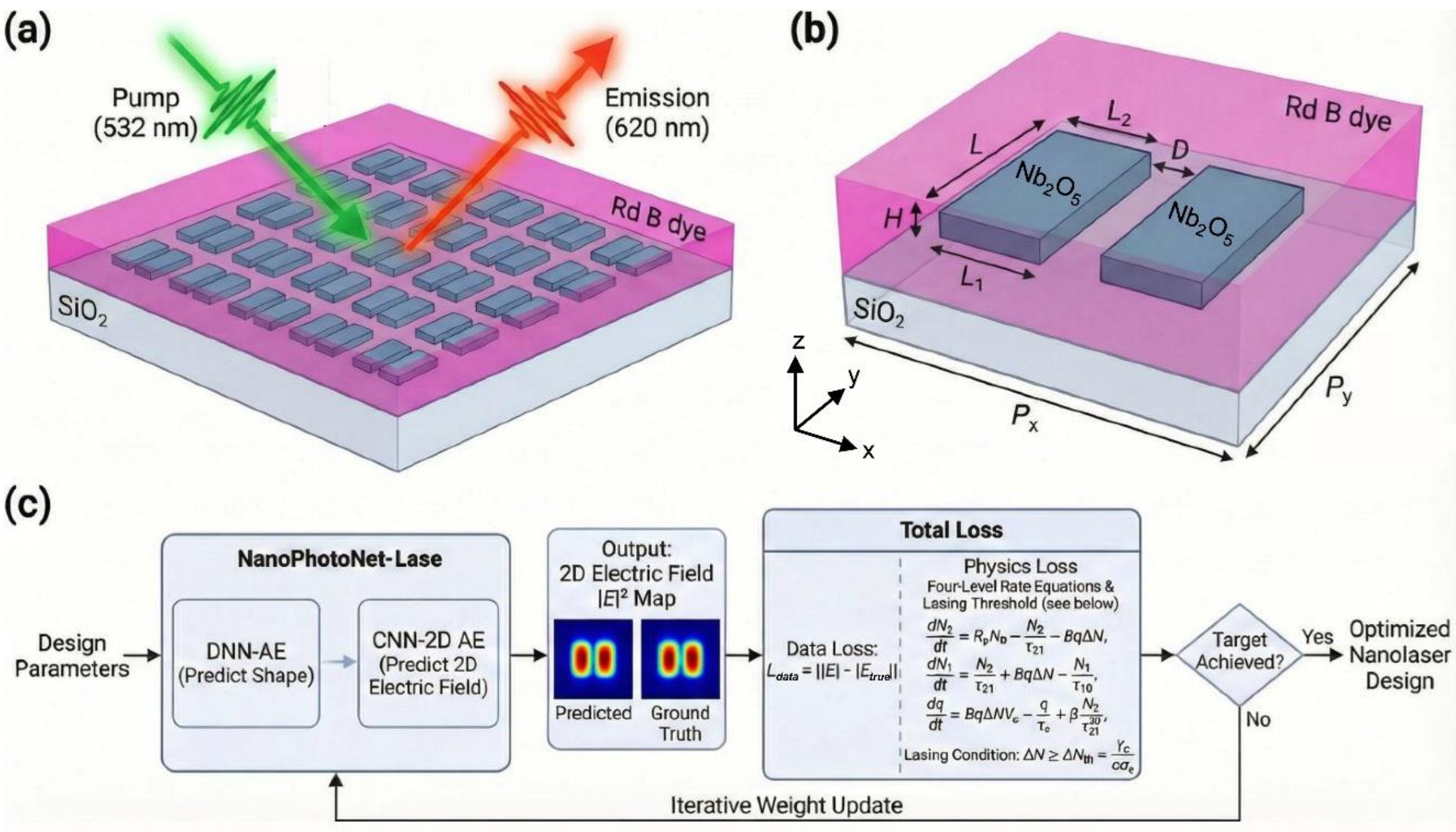

**Figure 1.** The proposed physics-informed neural network design for nanolasing from Rhodamine B (Rd B) dye, featuring a cavity that supports BIC resonance in the visible regime. (a) 3D schematic of the dielectric metasurface, consisting of two $Nb_2O_5$ pillars per meta-atom on a quartz substrate. It illustrates excitation by a green nanosecond pump laser (ns-laser) at a 532 nm wavelength and the resulting lasing emission from the Rd B dye. (b) Geometric parameters of the nanolaser meta-atom. The design parameters include the $x$-direction period ($P_x$), $y$-direction period ($P_y$), thickness of the $Nb_2O_5$ layer, a quartz substrate thickness of 500 µm, pillar length ($L$), top and bottom lengths ($L_1$ and $L_2$), and the horizontal separation between the two pillars ($D$). (c) Architecture of the NanoPhotoNet-Lase model, which consists of two coupled AI models to predict the optical response and the electric field of the light-matter interaction. The physics-informed loss function integrates rate equations and lasing conditions to assess the lasing threshold. Once the model is trained, iterative inverse design predictions are performed to achieve the target nanolaser design.

**Inverse Design and AI Prediction Performance**

Following structural definition and model initialization, NanoPhotoNet-Lase was deployed to inversely design the metasurface cavity parameters, such as in-plane periodicity, trapezoidal pillar dimensions, and inter-pillar gaps, which are required to achieve targeted lasing conditions. The model was optimized to locate high-$Q$ quasi-BIC modes through symmetry perturbations that intentionally introduce finite radiative leakage while preserving exceptional near-field localization.

The training evolution demonstrates exceptional convergence stability, rapidly minimizing the physics-constrained error to reach a training loss of merely 0.7% and a validation loss of 4.5% after 2000 epochs (Figure 2b). When tasked with inversely predicting a reflection spectrum with a target resonance near the desired emission wavelength of 620 nm, the AI model successfully generated optimal geometrical parameters that yielded a perfectly matched peak resonance (Figure 2a). The high-fidelity overlap between the FDTD-verified ground truth and the AI-predicted reflection profile highlights the robust capacity of the PINN to capture the complex, high-dimensional parameter space. Furthermore, the AI-predicted near-field electric field distribution exhibits a remarkably matched light-matter interaction at the nanoscale compared to the target profile, revealing strong localized enhancement within the dielectric pillars and excellent spatial overlap with the surrounding Rd B gain medium (Figure 2c).

To demonstrate the full versatility of the architecture, we extended its inverse design capabilities to encompass broadband single and dual resonances. In single resonance inverse design, the AI successfully reconstructed reflection spectra targeting specific $Q$-factors across a wide wavelength range from 620 nm to 700 nm (Section 4 in SI). More critically, the framework was tasked to inverse-design a high-$Q$ dual-resonant system capable of simultaneously coupling with both a specific pump wavelength and an emission wavelength. NanoPhotoNet-Lase accurately predicted configurations featuring tunable wavelength separations between the two resonances, ranging from 20 nm to 65 nm (Section 5 in SI). This dual-band capability is profoundly advantageous for accommodating diverse large-gain materials with varying Stokes shifts, solidifying the model's utility in executing complex nanophotonic design tasks with high computational efficiency.

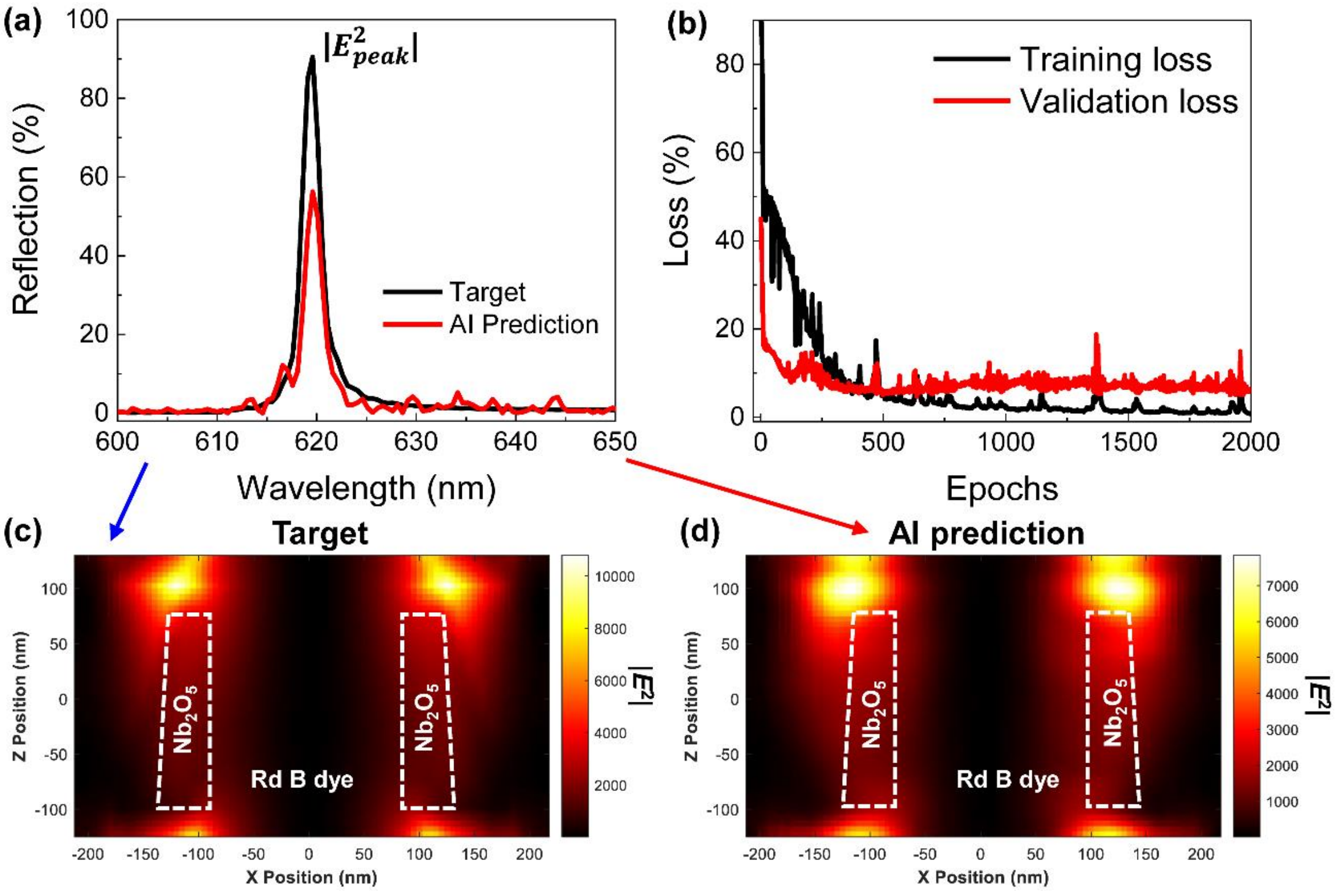

**Figure 2.** Prediction performance of the NanoPhotoNet-Lase model. (a) The target reflection spectrum with a resonance near the desired wavelength of 620 nm, plotted alongside the AI-predicted reflection spectrum from the inverse design, demonstrating a matched peak resonance wavelength. (b) The evolution of the NanoPhotoNet-Lase model's training over multiple epochs, achieving a training loss of 0.7% and a validation loss of 4.5%. (c) The target electric field profile at the resonant wavelength of 620 nm. (d) The AI-predicted electric field profile, showing an accurate match for the nanoscale light-matter interaction.

**Experimental Realization and Nanolasing Characteristics**

To rigorously validate the computational predictions generated by NanoPhotoNet-Lase, we fabricated the inversely designed $Nb_2O_5$ dielectric metasurface cavities and characterized their active nanolasing performance. The nanofabrication workflow relied on a combination of precision electron-beam lithography (EBL) and atomic layer deposition (ALD) to ensure high-fidelity replication of the AI-derived geometrical parameters. Inductively coupled plasma reactive ion etching (ICP-RIE) utilizing $CHF_3$ chemistry was employed to define the sharp trapezoidal pillars, followed by an $O_2$ plasma descum to remove the residual resist. SEM characterization confirms excellent dimensional agreement with the targeted AI predictions (Figure 3a), revealing well-defined, uniform meta-atoms with the required inter-pillar gaps crucial for quasi-BIC excitation. Finally, the organic gain layer was spin-coated onto the sample to form the active, index-symmetric cavity (Figure S5).

The optical properties and lasing behaviors were characterized using a custom micro-photoluminescence setup (Figure S6). The measured normalized PL intensity as a function of the incident pump fluence demonstrates a clear transition from spontaneous emission to amplified stimulated emission. At low pump fluences, a broad, low-intensity spontaneous emission spectrum is observed. As the fluence incrementally increases, a distinct, intense lasing peak violently emerges near the target wavelength of 620 nm (Figure 3b). A highly nonlinear "S-curve" response is evident when plotting the peak PL intensity against the pump fluence, explicitly defining the onset of the lasing threshold at an extraordinarily low $P_{th} \sim 565\ \mu J/cm^2$ (Figure 3c). Concurrently, we observe a dramatic collapse of the emission linewidth; the measured full width at half maximum (FWHM) rapidly narrows from over 50 nm down to less than 4 nm at fluences strictly exceeding the threshold (Figure 3d).

Crucially, the experimental data exhibits remarkable quantitative alignment with the NanoPhotoNet-Lase predictions. The measured threshold fluence and the specific emission wavelength (simulated 620 nm) display less than a 1% deviation from the physical constraints learned by the AI model. Further confirmation of spatial coherence is verified through far-field optical imaging. At pump fluences below the threshold, the filtered PL emission manifests as a weak, diffuse glow distributed isotropically across the field of view (Figure 3e). Conversely, immediately above the lasing threshold, the emission transforms into a highly intense, bright, and localized signal confined strictly within the patterned metasurface boundary (Figure 3f), confirming the successful activation of the high-$Q$ quasi-BIC nanolaser array precisely engineered by the physics-informed AI.

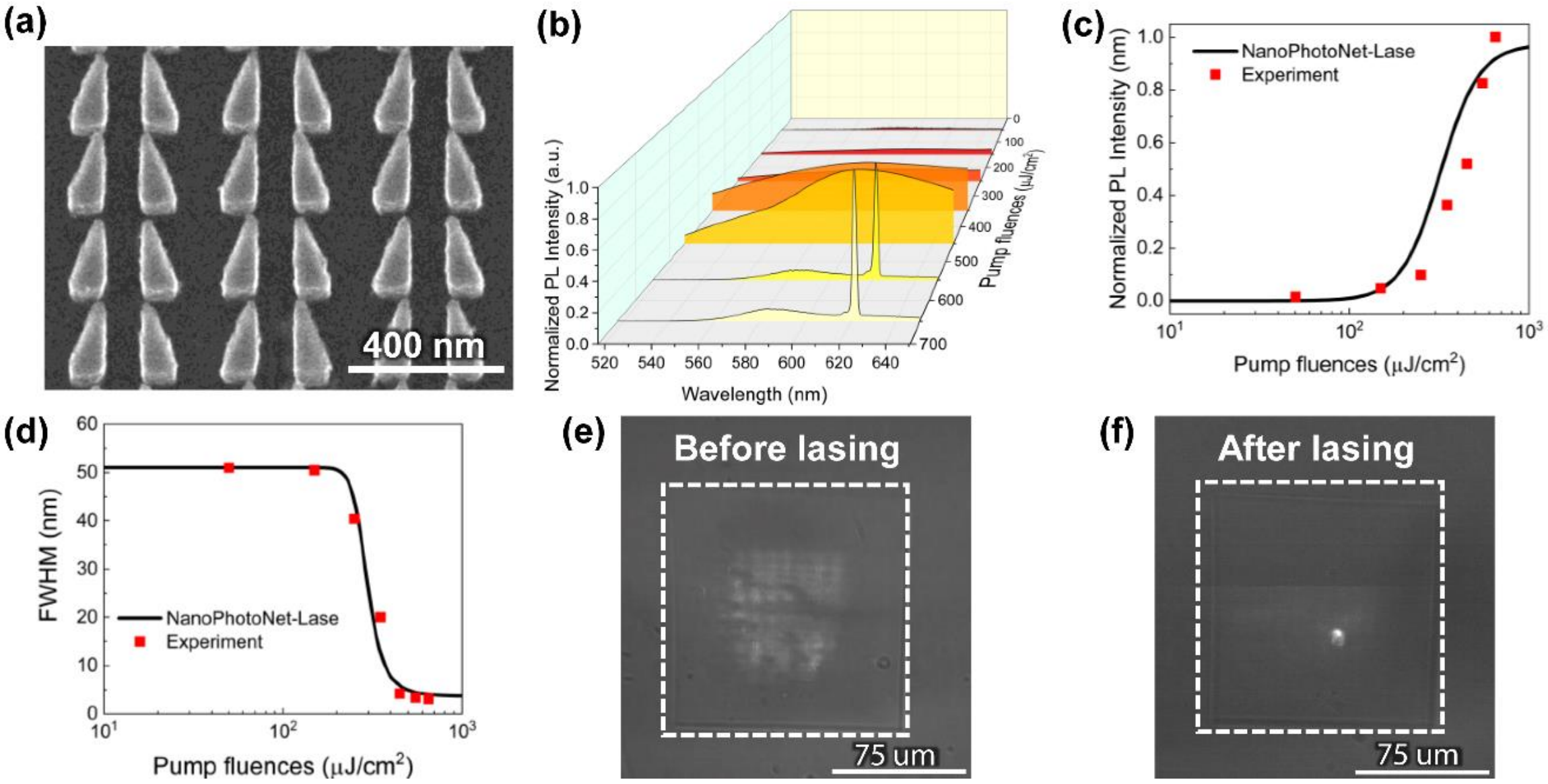

**Figure 3.** Experimental validation of the NanoPhotoNet-Lase model's predictions. (a) Scanning electron microscope (SEM) image of the fabricated dielectric metasurface cavity, featuring dimensions inversely predicted by the AI model. (b) Measured normalized photoluminescence (PL) as a function of pump fluence, illustrating the emergence of a lasing emission peak near the 620 nm wavelength. (c) Comparison between the measured normalized peak PL versus pump fluence and the PL emissions predicted by NanoPhotoNet-Lase. (d) Measured full width at half maximum (FWHM) of the PL emission versus pump fluence, plotted alongside the FWHM predicted by the model. (e) Optical image of the filtered PL emission at a pump fluence below the lasing threshold, showing low emission intensity. (f) Optical image of the filtered PL emission at a pump fluence above the lasing threshold, displaying strong, bright emission intensity within the metasurface area.

## Coherent Wavefront Engineering: Phase-Gradient Nanolaser and Beam Steering

Beyond uniform emission, modern active nanophotonics demands the capacity to spatiotemporally manipulate generated coherent light. Relying upon the fundamental principle of optical reciprocity, the far-field emission profile of quantum emitters embedded within a resonant cavity directly mirrors the localized near-field response excited under external illumination.[39] To showcase the profound scalability and programmability of the NanoPhotoNet-Lase model, we extended the inverse design framework to engineer phase-gradient metasurfaces capable of beam focusing and directional steering through predicting the phase of transmitted electric field. Then, applying the Pancharatnam-Berry (PB) phase concept to spatially varying phase shifts across the metasurface array by rotating individual trapezoidal meta-atoms relative to the $z$-axis. Comprehensive transmission analysis confirms that adjusting the rotation angle achieves a complete $2\pi$ spatial phase span while maintaining an exceptional transmission efficiency approaching 98% near the resonant lasing emission wavelength (Figure S7).

Leveraging this precise phase control, we first designed a light-emitting metalens to focus the amplified nanolaser emission (Figure 4a). The spatial phase distribution across the metasurface (Figure 4b) was programmed to follow an ideal hyperbolic phase profile (Equation 18 in SI). The AI-optimized unit cell rotation map accurately reconstructs this phase gradient. Three-dimensional FDTD simulations of the active metalens reveal a highly focused, diffraction-limited far-field spot located precisely at the focal plane (Figure 4c), characterized by a minimal FWHM of approximately 391 nm. While minor side lobes are present due to subtle residual scattering and phase discretization errors, the overarching focusing efficiency validates the successful integration of gain dynamics with complex wavefront manipulation.

Furthermore, we demonstrated coherent beam steering by imposing a strict linear phase gradient across the active metasurface array, defined by a specific linear phase distribution (Equation 19 in SI). Targeting steering angle of 15 degrees (Figure 4d), the resulting meta-atom rotation map smoothly transitions the optical phase along the $x$-axis (Figure 4e). Simulated far-field propagation of the nanolaser explicitly confirms that the intense, coherent emission is successfully redirected into a unidirectional plane wave tilted precisely at the target angle (Figure 4f), with minimal crosstalk into secondary diffraction orders. The successful realization of both focusing and steering functionalities underscores the immense potential of integrating physics-informed deep learning with the PB phase paradigm to develop fully programmable, on-chip coherent light sources.

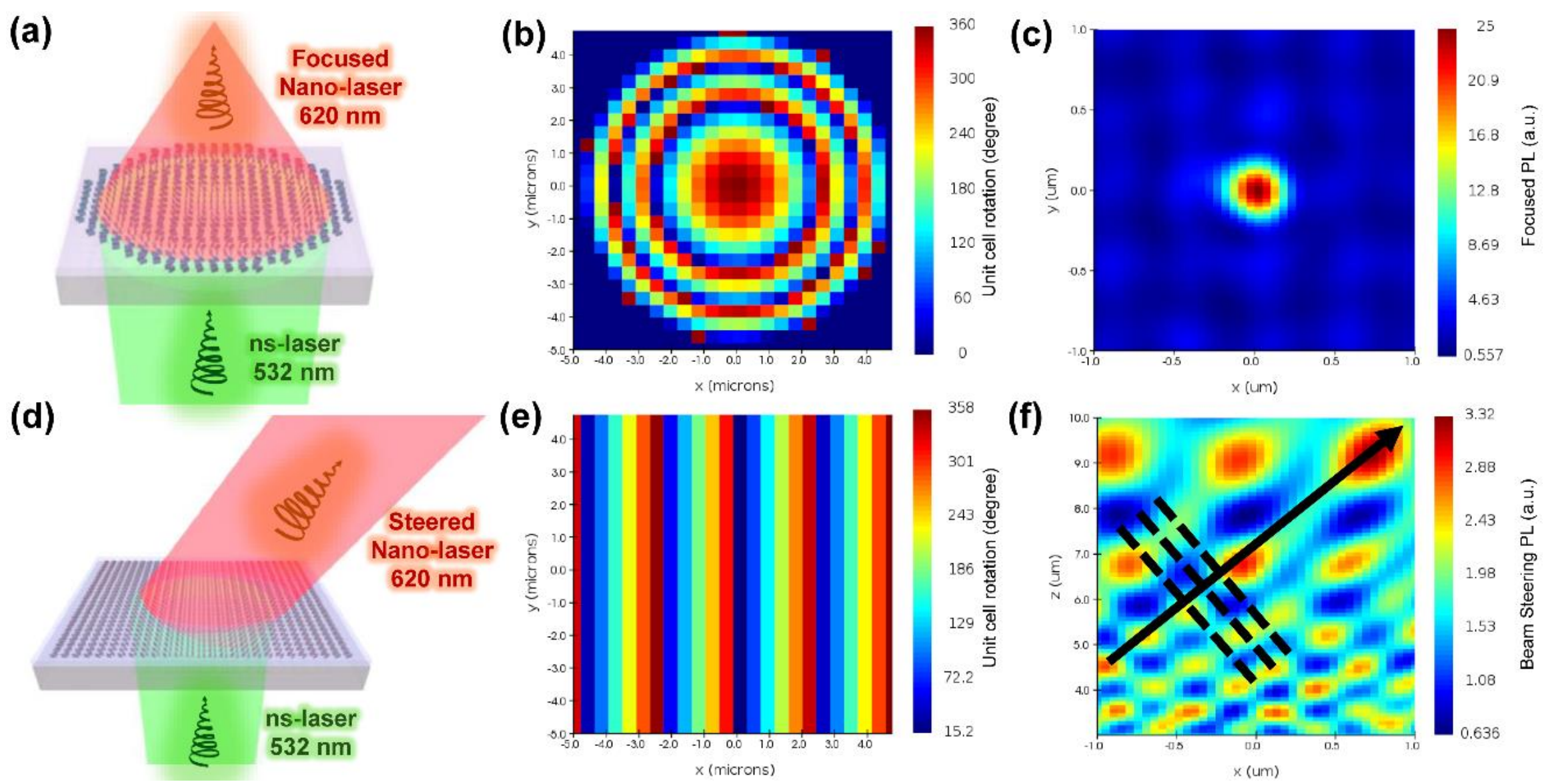

**Figure 4.** Design of a phase-gradient nanolaser metasurface. (a) Schematic of the proposed focused nanolaser emission, utilizing a metalens designed based on the Pancharatnam-Berry (PB) phase concept. (b) Meta-atom rotation map and the corresponding phase map ranging from zero to $2\pi$. (c) Simulated focused nanolaser emission intensity, demonstrating a focal spot with an FWHM of ~391 nm. (d) Schematic of the proposed steered nanolaser emission, using a beam-steering metasurface based on the PB phase concept. (e) Meta-atom rotation map designed to steer light by 15 degrees. (f) Simulated far-field steered nanolaser emission, showing plane waves steered at an angle of 15 degrees.

To contextualize the scientific advancement introduced by our model, we conducted a rigorous benchmark of NanoPhotoNet-Lase against existing AI and PINN frameworks found in current literature. Recent applications of deep learning in nanophotonics have largely relied on purely data-driven architectures to predict passive dielectric properties or structural colors.[40] However, these conventional frameworks explicitly omit the physical processes that govern stimulated emission, gain depletion, and carrier population dynamics, rendering them functionally incapable of designing active, lasing systems. While standard PINNs have been successfully implemented to solve Maxwell's equations for passive 3D light diffraction,[35, 41] NanoPhotoNet-Lase represents a significant paradigm shift by simultaneously embedding the vector Helmholtz equation and the complex four-level laser rate equations into a singular, unified optimization framework. This tightly coupled formulation permits the self-consistent prediction of both the electromagnetic cavity response and the nonlinear gain dynamics, allowing the network to directly calculate non-trivial lasing thresholds without an exhaustive reliance on pre-labeled simulation databases.

As highlighted in the benchmark (Table S1), NanoPhotoNet-Lase optimized a quasi-BIC resonance yielding an exceptional $Q$-factor of 264, while identically optimizing the required PL amplification and the corresponding phase-gradient functionality. Our framework not only facilitates single-wavelength optimization but supports a broad dimensional tuning range from 620 nm to 695 nm, a capability largely absent from passive AI models. The simultaneous achievement of ultra-low threshold coherent lasing, dynamic tunability, and precise far-field wavefront control proves that NanoPhotoNet-Lase extends the utility of machine learning in photonics far beyond passive scattering problems. By grounding the neural network explicitly in the physical laws of optical gain and electromagnetism, this methodology establishes a highly efficient, scalable, and inherently interpretable pathway for the inverse design of the next generation of active, programmable optoelectronic devices.

## CONCLUSION

We have demonstrated NanoPhotoNet-Lase, a physics-informed learning framework that unifies laser rate-equation physics with metasurface electromagnetics, achieving real-time prediction of nanolaser performance with fidelity approaching numerical solvers but orders-of-magnitude faster computation. Using $Nb_2O_5$-RhB metasurface arrays, we validated the model through direct fabrication and lasing experiments, achieving quantitative agreement in wavelength and threshold. Moreover, the approach enabled phase-gradient nanolasers and beam steering exhibiting coherent beam focusing and steering respectively which considered a fundamental step toward active, programmable metasurface emitters. This work pioneers the integration of physics-informed artificial intelligence in active nanophotonics, demonstrating how embedded physical laws can overcome data scarcity and ensure physically realistic predictions. Looking forward, the differentiable architecture of NanoPhotoNet-Lase could be extended to gradient-based inverse design, electrically injected systems, or adaptive metasurfaces incorporating phase-change materials for real-time beam control. Furthermore, this framework provides a generalizable methodology for physics-governed AI acceleration in light–matter systems of several optoelectronic devices.[42-59]

## METHODS

### Reflection and Field Profile Simulations

A three-dimensional finite-difference time-domain (FDTD) commercial solver (Lumerical FDTD Solutions) was used to perform parametric sweeps over the geometry and refractive indices of the BIC metasurfaces.[60] The complex refractive index dispersion of $Nb_2O_5$ was taken from ellipsometry measurements of the corresponding thin films (Figure S2 in SI). The active metasurface cavity was illuminated by a normally incident plane wave with linear polarization along the x-axis. Periodic boundary conditions were applied along the in-plane ($x$ and $y$) directions to model an effectively infinite array, while perfectly matched layers were used along the z-direction to absorb outgoing radiation and suppress artificial reflections. A nonuniform mesh with a minimum step size below 5 nm was employed in the vicinity of the nanostructures to accurately resolve their features. Transmission spectra were recorded using a frequency-domain monitor placed in the quartz substrate, and spatial field distributions at the BIC resonances were extracted from two-dimensional monitors intersecting the $Nb_2O_5$ polygonal pillars. A stringent automatic shutoff level of $1\times10^{-7}$ was enforced to ensure convergence to a steady-state solution. For each geometry in the sweep, the transmission, reflection, and full three-dimensional electric-field tensors were exported for subsequent analysis and for training the NanoPhotoNet-Lase model.

**Fabrication of all-optical BIC nanolaser metasurfaces**

The fabrication workflow for the tunable hybrid metasurface is summarized in (Figure S5). Double-side polished deep-UV-grade quartz substrates (Photonik, Singapore) were first cleaned by sequential 10 min ultrasonic baths in acetone and isopropanol (IPA), followed by rinsing and nitrogen blow-drying. Nanopillar molds for the $Nb_2O_5$ metasurface were defined by electron-beam lithography (EBL, Elionix ELS-7000). A positive-tone resist (ZEP520A) was spin-coated at 3000 rpm for 90 sec to yield a thickness of approximately 400 nm and baked at 180 °C for 2 min. To mitigate charging on the insulating substrate, a conductive polymer layer (Espacer) was spin-coated at 1500 rpm for 30 s prior to exposure. Arrays with lateral dimensions of $300\times300\ \mu m^2$ were written at 100 kV accelerating voltage, using a 200 pA beam current and a nominal dose of 320 μC/cm$^2$. After exposure, the Espacer layer was removed in deionized water, and the resist was developed in xylene for 30 s and quenched in IPA. A 100 nm-thick $Nb_2O_5$ film was then deposited by atomic layer deposition (ALD, Beneq TFS 200) at 85 °C, below the glass transition temperature of ZEP520A. The deposition employed tert-butylimino tris(diethylamido) niobium as the metal precursor and water as the oxidant, resulting in a growth rate of approximately 0.59 Å per cycle. To remove the overfilled $Nb_2O_5$ layer on top of the resist and define the pillars, a blanket etch was performed using inductively coupled plasma reactive ion etching (ICP-RIE, Oxford OIPT Plasmalab). The etch used a $CHF_3$ plasma (25 sccm) at 150 W RF power, 900 W ICP power, and 25 mTorr pressure, corresponding to an etch rate of roughly 150 nm/min. Residual ZEP520A was subsequently stripped by an oxygen plasma descum (50 sccm $O_2$, 250 W RF, 3 min). To promote adhesion and infiltration of the gain medium, the metasurface was exposed to UV radiation under vacuum for 10 min immediately before coating. Finally, the organic gain layer, consisting of RhB dissolved in a polyvinyl alcohol (PVA) matrix at the desired concentration, was spin-coated onto the metasurface to form the active, index-symmetric cavity.

**Transmission measurements**

Far-field transmission spectra of the metasurfaces were measured using a supercontinuum nanosecond laser source (Leukos Opera, 30 kHz repetition rate). The incident beam was conditioned to be linearly polarized along the *x*-direction, aligned with the dominant resonance, by passing through a sequence of polarization optics comprising a quarter-wave plate, a half-wave plate, and a linear polarizer. Light transmitted through the sample was collected by an optical fiber positioned behind the substrate and analyzed using an Ocean Optics USB4000 spectrometer.

**Nanolasing measurements**

Lasing performance was characterized using a micro-photoluminescence (μ-PL) configuration, schematically illustrated in (Figure S6**)**. The metasurface was pumped by a 532 nm nanosecond laser operating in free space (10 Hz repetition rate, average power up to 50 mW, pulse energies up to 150 μJ). The pump fluence was adjusted using a variable optical attenuator. A linear polarizer and half-wave plate were used to set the pump polarization to match the resonant mode of the metasurface. The beam was focused onto the sample using a focusing lens, ensuring uniform excitation over the patterned region. The sample was mounted on a motorized XYZ stage to enable fine control of lateral position and focus. Emission from the metasurface was collected by a second lens and passed through a 532 nm notch filter to suppress residual pump light. For sample alignment and real-space imaging, a white-light source, beam splitter (glass plate), and CCD camera were integrated into the optical path via a flip mirror (M1) and imaging lens (L1). Lasing spectra were recorded using a WiTEC photoluminescence spectrometer equipped with gratings of 150 or 600 grooves/mm, depending on the required spectral resolution.

**ASSOCIATED CONTENT**

**Supporting Information**

The theoretical framework of nanolaser physics; measured ellipsometer refractive indices of $Nb_2O_5$; S4. NanoPhotoNet-Lase: Single resonance inverse design; NanoPhotoNet-Lase: Dual resonance inverse design; nanofabrication flow for the nanolaser metasurface cavity;

Nanolaser measurement optical setup; Phase versus PINNs nanolaser meta-atom rotation; PINNs nanolaser metalens; PINNs nanolaser beam steering; Benchmark of NanoPhotoNet-Lase with AI and PINN Design Models.

**CORRESPONDING AUTHORS**

Email: Omar_Abdelrahman@a-star.edu.sg

**ORCID**

Omar A. M. Abdelraouf: https://orcid.org/0000-0002-9065-7414

**AUTHOR CONTRIBUTIONS**

O.A.M.A. conceived the physical concept of PINN for nanolasers, idea of NanoPhotoNet-lase model, designed the active BIC nanolaser cavity, built dataset using 3D FDTD simulations, and did materials deposition and optical characterization, cleanroom nanofabrication of optimized BIC nanolaser cavity, linear optical measurements, photoluminescence and lasing measurements, design phase gradient nanolaser, and wrote the whole manuscript.

**Competing Financial Interests**

The authors declare no competing financial interest.

**ACKNOWLEDGMENT**

Authors acknowledge funding from A*STAR under its Career Development Fund (CDF) grant no. C233312016 and SERC Central Research Funds (CRF). Also, the support from Singapore international graduate award (SINGA).